\documentclass[sigconf]{acmart}

\AtBeginDocument{%
  }

\usepackage{booktabs}
\usepackage{xcolor}
\usepackage{colortbl}
\definecolor{bestcolor}{RGB}{255,179,71}
\definecolor{secondcolor}{RGB}{255,255,153}

\begin{document}

\title{From Item-Only to Query-Item: Query-Conditioned Generative Search with QGS in Quark}

\author{Yanglong Song}
\email{songyanglong.syl@alibaba-inc.com}
\affiliation{%
  \institution{Alibaba Group}
  \country{China}
}
\affiliation{%
  \institution{University of Science and Technology of China}
  \city{Hefei}
  \country{China}
}

\author{Zihao Yang}
\email{yuzhibo.yzh@alibaba-inc.com}
\affiliation{%
  \institution{Alibaba Group}
  \country{China}
}
\affiliation{%
  \institution{University of Science and Technology of China}
  \city{Hefei}
  \country{China}
}

\author{Shuo Meng\textsuperscript{$\dagger$}}
\email{mengshuo.meng@alibaba-inc.com}
\affiliation{%
  \institution{Alibaba Group}
  \country{China}
}

\author{Rujun Guo}
\email{rujun.grj@alibaba-inc.com}
\affiliation{%
  \institution{Alibaba Group}
  \country{China}
}

\author{Jin Zhang}
\email{stark.zj@alibaba-inc.com}
\affiliation{%
  \institution{Alibaba Group}
  \country{China}
}

\author{Bin Wang}
\email{yaoming.wb@alibaba-inc.com}
\affiliation{%
  \institution{Alibaba Group}
  \country{China}
}

\author{Shaoyu Liu}
\email{liushaoyu.lsy@alibaba-inc.com}
\affiliation{%
  \institution{Alibaba Group}
  \country{China}
}

\author{Xiaozhao Wang}
\email{orlando@alibaba-inc.com}
\affiliation{%
  \institution{Alibaba Group}
  \country{China}
}

\author{Guanjun Jiang}
\email{guanj.jianggj@alibaba-inc.com}
\affiliation{%
  \institution{Alibaba Group}
  \country{China}
}

\renewcommand{\shortauthors}{Song et al.}
\begin{abstract}
Generative sequence models have shown strong results in recommendation. Applying them to search ranking is more challenging. Search behavior is inherently query-driven. Each query switch introduces a sharp topic shift in the user's interaction history. Existing generative methods flatten queries and items into a single token sequence. They do not distinguish query boundaries. This causes the model to mix different query intents into one prediction target, resulting in noisy supervision.

We present Query-Conditioned Generative Search (QGS). QGS encodes each interaction as a $(query, item)$ pair token. It trains with a query-conditioned next-item objective. The prediction target changes from a noisy marginal $P(\mathrm{item}_{t+1}\mid \mathrm{context}_{\leq t})$ to a clean conditional $P(\mathrm{item}_{t+1}\mid \mathrm{context}_{\leq t}, \mathrm{query}_{t+1})$. This directly removes the semantic discontinuity caused by query switches.

Encoding long interaction histories with standard attention has quadratic cost. This is impractical under strict online latency budgets. We introduce a Linear HSTU encoder. It replaces full attention with causal linear recurrence. Per-layer complexity drops from $O(L^2)$ to $O(L)$ with no loss in ranking quality.

Traditional search ranking depends on hand-crafted features like text-matching scores, statistical signals, and behavioral features. We propose HFG-Attention to preserve them in the generative framework. It organizes heterogeneous features into semantic groups and fuses them through a dedicated attention block. This bridges sparse engineered signals with dense sequential representations.

QGS is deployed in the ranking module of Quark Search, a major commercial search engine in China. Online A/B tests show statistically significant gains: +0.62\% CTR, +0.38\% Click-Search Ratio, and +3.55\% PV Duration over the production deep learning baseline.
\end{abstract}


\maketitle
\renewcommand{\thefootnote}{\fnsymbol{footnote}}
\footnotetext[2]{Corresponding author.}
\renewcommand{\thefootnote}{\arabic{footnote}}

\section{Introduction}

\begin{figure*}[t]
  \centering
  \includegraphics[width=0.65\textwidth]{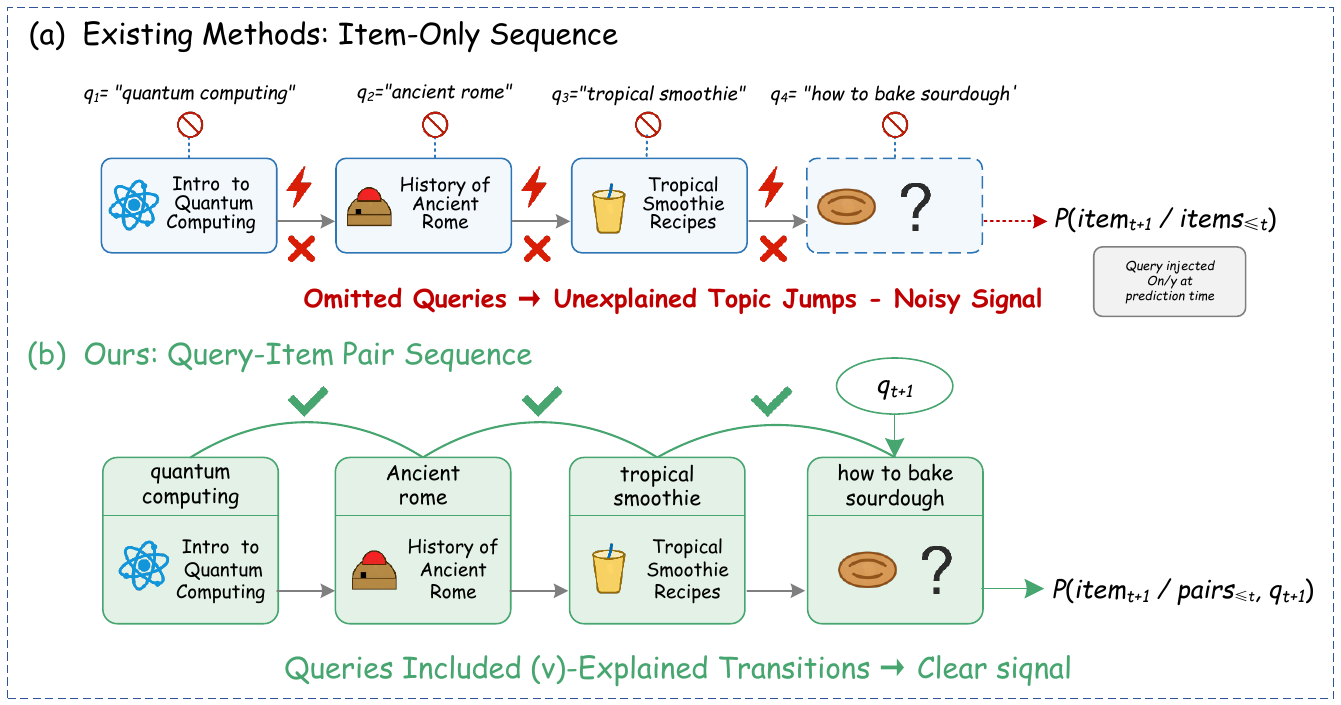}
  \caption{Overview of QGS, a generative framework for search. Unlike prior generative recommendation/search methods that model item-only sequences, QGS autoregressively models query-conditioned next-item prediction, explicitly capturing the query-conditioned nature of search behavior.}
  \Description{A schematic comparison between prior item-only generative sequence modeling and the proposed query-conditioned generative search framework, showing how the next query is fused with the historical user representation to predict the next clicked document.}
  \label{fig:teaser}
\end{figure*}

Search is a core component of modern information retrieval systems, serving billions of users daily across e-commerce, short-video, and web search scenarios. Given a user query, a search system must retrieve and rank the most relevant documents from a massive candidate pool. Industrial search systems typically adopt a multi-stage cascading architecture, sequentially proceeding through recall, pre-ranking, and ranking, where the ranking stage relies on deep ranking models such as Wide\&Deep~\cite{Cheng2016WideD}, DeepFM~\cite{Guo2017DeepFMAF}, and DIN~\cite{Zhou2017DeepIN}. These models score candidate documents based on a large number of hand-crafted features---text-matching features, statistical features, and contextual features---and have become the dominant paradigm for industrial search ranking.

In recent years, generative sequence modeling has achieved remarkable success in recommendation. It formulates user behavior history as autoregressive next-item prediction. Inspired by this success, researchers have naturally attempted to transfer this paradigm to search. However, the straightforward transfer shares a common problem: none of the existing methods explicitly models the conditional dependency between queries and items during autoregression.

This causes problems in search. Search sequences are fundamentally different from recommendation sequences. Every new query the user issues triggers an entirely fresh candidate set. Adjacent items can undergo drastic topical jumps---for example, from \textit{quantum computing} to \textit{ancient Roman history} to \textit{tropical smoothies}. In an item-only sequence, these jumps appear as inexplicable semantic discontinuities, making the next-item prediction signal extremely noisy. Yet these seemingly random transitions are entirely driven by query switches. \textbf{Every item has a corresponding query as its causal explanation. Therefore, in search, the query is not optional auxiliary information but a necessary condition for eliminating sequence-level semantic discontinuity and turning the prediction signal from noisy into clean}.

To address this gap, we propose \textbf{Query-Conditioned Generative Search (QGS)}, a generative framework for search ranking. QGS performs Query-Item Pair Token Construction at each step of the user interaction sequence. It concatenates the textual semantic embeddings of both the query and the item with numerical and categorical feature embeddings into a unified pair token representation. Building upon this representation, QGS adopts \textbf{Query-Conditioned Next-Item Prediction} as its core training objective. At each step $t$, the model conditions on both the historical context and the query at step $t{+}1$ to predict the item interacted with at step $t{+}1$. The prediction is optimized with an InfoNCE~\cite{oord2018representation} contrastive loss. This reformulates next-item prediction from the noisy marginal $P(\mathrm{item}_{t+1}\mid \mathrm{context}_{\leq t})$ into the clean conditional $P(\mathrm{item}_{t+1}\mid \mathrm{context}_{\leq t}, \mathrm{query}_{t+1})$. The item jumps caused by query switches now acquire a clear causal explanation. Furthermore, the query-conditioned objective requires encoding long user interaction histories. Standard attention incurs $O(L^2)$ cost, which is prohibitive for online serving. We therefore propose a Linear HSTU Encoder. It integrates HSTU's U-gating mechanism with the linear attention paradigm of RWKV~\cite{peng2023rwkv}, reducing complexity to $O(L)$. At sequence length $1{,}000$, the encoder achieves a $37\%$ inference speedup with no degradation in ranking quality. Finally, the candidate item is the prediction target and cannot appear in the sequence input. However, its cross features with the user remain critical for ranking quality. To this end, we introduce Heterogeneous Feature Grouping Attention (HFG-Attention). It organizes heterogeneous sparse features into semantic groups and fuses them with dense sequential representations via an HSTU attention block.

We conduct extensive experiments on a large-scale industrial search ranking system. QGS achieves consistent and significant improvements across multiple offline and online metrics.

The main contributions of this paper are summarized as follows:
\begin{itemize}
    \item \textbf{Query-Conditioned Next-Item Prediction.} We propose query-conditioned next-item prediction as the core training objective of QGS. At each autoregressive step, the model explicitly conditions on the next query to predict the next item. This transforms the noisy marginal prediction in search sequences into a clean conditional prediction. Query-Item Pair Token Construction serves as the input representation layer, ensuring that every step carries full query-item interaction information. This design directly remedies the missing query-item conditional dependency in existing generative search methods.
    \item \textbf{Linear HSTU Encoder.} We integrate HSTU's U-gating with the linear attention paradigm of RWKV, reducing complexity from $O(L^2)$ to $O(L)$. At sequence length $1{,}000$, the encoder achieves a $37\%$ inference speedup while matching the original HSTU in ranking quality.
    \item \textbf{HFG-Attention.} We organize heterogeneous sparse contextual features into semantic groups and fuse them with dense sequential representations via an HSTU attention block with the FFN retained, resolving the heterogeneous feature fusion challenge in generative ranking.
    \item \textbf{Industrial-scale Empirical Validation.} We deploy QGS on Quark Search, with consistent and significant improvements across multiple offline and online metrics.
\end{itemize}

\section{Related Works}

\subsection{Deep Ranking Models and Feature Interaction}

Industrial search ranking is dominated by discriminative deep models. Wide\&Deep~\cite{Cheng2016WideD} combines a linear model with a deep network; DeepFM~\cite{Guo2017DeepFMAF} adds factorization machines for automatic low-order feature crosses; DIN~\cite{Zhou2017DeepIN} uses target attention to capture query-relevant historical interests; SIM~\cite{pi2020search} extends behavior modeling to lifelong sequences via search-based retrieval. These models typically adopt a dual-pathway architecture where numerical features, categorical embeddings, and text semantic features from pre-trained models such as BERT~\cite{devlin2019bert} are fused before multi-task prediction towers. A parallel line of work focuses on efficient feature interaction: DCN~\cite{wang2017deep} introduces explicit bounded-degree feature crossing; FiBiNET~\cite{huang2019fibinet} dynamically learns feature importance via Squeeze-and-Excitation; AutoInt~\cite{song2019autoint} uses multi-head self-attention for high-order interactions; DCN V2~\cite{wang2021dcn} optimizes cross networks for large-scale sparse settings; RankMixer~\cite{zhu2025rankmixer} proposes hardware-aware token mixing with per-token FFN. Our HFG-Attention borrows the \textit{token-level attention interaction} idea from this line, tailoring it for the heterogeneous-fusion semantic gap unique to generative ranking.

\subsection{Generative Sequence Modeling for Recommendation and Search}

Generative sequence modeling reformulates retrieval as autoregressive next-item prediction. In recommendation, SASRec~\cite{kang2018self} establishes the Transformer-based sequential foundation. TIGER~\cite{rajput2023recommender} assigns Semantic IDs via RQ-VAE for generative retrieval. LC-Rec~\cite{zheng2024adapting} integrates collaborative and language semantics through learned item indices. HSTU~\cite{Zhai2024ActionsSL} scales to trillions of parameters with softmax-free attention and U-gating. HLLM~\cite{Chen2024HLLMES} introduces a hierarchical LLM architecture with separate Item and User LLMs. OneRec~\cite{Deng2025OneRecUR} unifies recommendation as generic sequence prediction. Inspired by this success, several works extend generative modeling to search. They differ significantly in task definition and how they handle query information, and can be categorized into two groups.

\textbf{(1) Single-query Generative Retrieval.} These methods define search as ``given a query, autoregressively generate the target item's Semantic ID.'' TSG~\cite{zhang2024generative} and QueStER~\cite{satouf2026quester} both belong to this category. Given a user query, the model autoregressively generates the multi-level codes of the target item's Semantic ID (learned via RQ-VAE or similar). The tokens in their autoregressive sequence are \textit{item ID codes}, not multiple user interactions. The query serves as an input condition for generation but does not appear inside the autoregressive sequence. GBS~\cite{tang2025generative} adopts a similar paradigm in the book search domain with data augmentation. These methods are essentially generative versions of single query-to-item mapping. They do not model user behavior evolution across queries.

\textbf{(2) User-sequence Generative Search.} These methods model search as autoregressive sequence prediction over user interaction histories. UniSearch~\cite{chen2025unisearch} and OneSearch~\cite{chen2025onesearch} unify search into a generative framework and autoregressively predict the next interacted item on user behavior sequences. However, their sequences are still dominated by item tokens. Queries are only injected as prefix conditions or external inputs. The query and item are not explicitly conditioned upon each other during autoregression. GenSAR~\cite{10.1145/3705328.3748071} unifies generative modeling for search and recommendation. It models the interaction prediction among search results, queries, and recommendation results, but does not use the query as a condition at each autoregressive step. LORE~\cite{lu2025lore} leverages large language models for query-document relevance judgment and does not follow a sequence modeling paradigm.

Overall, none of the existing generative search methods explicitly perform query-conditioned next-item prediction at the training objective level---i.e., explicitly conditioning the next-item prediction on the corresponding \textit{next} query at each autoregressive step. Our method fills this gap by conditioning every autoregressive step on the corresponding next query, directly modeling that each item click is driven by its query.

\subsection{Linear Attention}

The $O(L^2)$ complexity of vanilla Transformers~\cite{vaswani2017attention} becomes prohibitive for long behavior sequences in online serving. Linear Transformer~\cite{katharopoulos2020transformers} approximates softmax attention via kernel feature maps but may accumulate approximation error; RWKV~\cite{peng2023rwkv} reformulates attention as a linear recurrence with element-wise key-value accumulation; RetNet~\cite{sun2023retentive} introduces the Retention mechanism supporting parallel, recurrent, and chunked modes; Mamba~\cite{gu2023mamba} employs selective state space models with input-dependent parameterization. However, directly applying these methods to HSTU would discard its softmax-free design and U-gate---the core sources of HSTU's success. Our Linear HSTU encoder borrows the linear-recurrence form of RWKV while fully retaining HSTU's softmax-free design and U-gating, reducing complexity to $O(L)$ without sacrificing modeling capacity.
\begin{figure*}[t]
    \centering
    \includegraphics[width=0.85\textwidth]{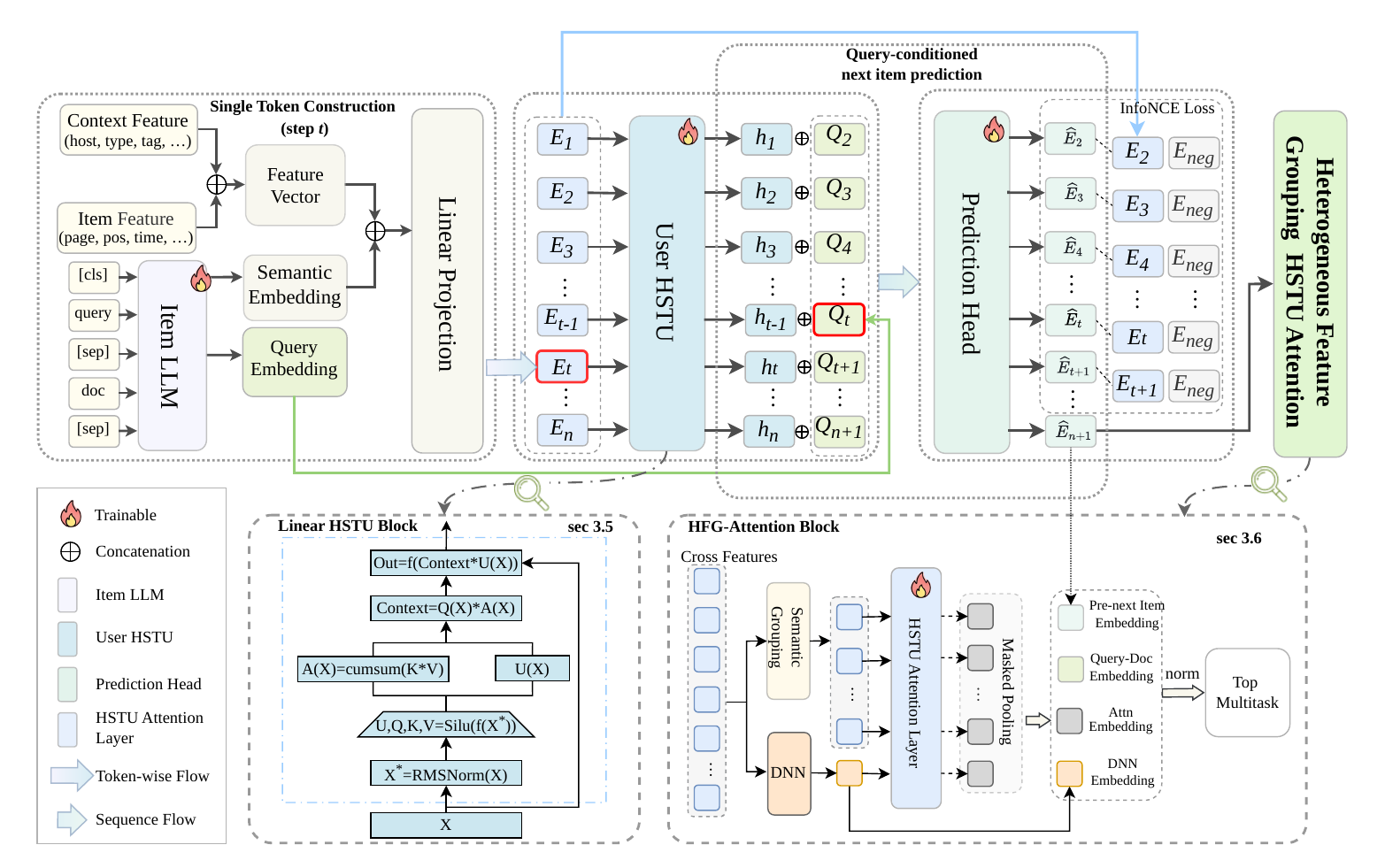}
    \caption{Overall framework of QGS. \textbf{(Top)} Each search interaction is encoded as a query-item pair token via an Item LLM and feature projection, then sequentially modeled by the Linear HSTU encoder; the encoder output is concatenated with the next query and scored against in-batch negatives via an InfoNCE loss. \textbf{(Bottom-left)} Internal structure of a Linear HSTU block (Section~\ref{sec:linear_hstu}), which reduces per-layer complexity from $O(L^2)$ to $O(L)$. \textbf{(Bottom-right)} HFG-Attention block (Section~\ref{sec:sparse_attn}), which fuses sparse cross features with dense sequential representations for the final multi-task prediction.}
    \label{fig:framework}
\end{figure*}
\section{Method}

\subsection{Problem Definition and Data Construction}
\label{sec:problem}

\paragraph{Task Definition.}
Given a user's historical search interaction sequence and a candidate query-document pair, our task is to predict the probability that the user clicks the candidate document under the current query, thereby producing an accurate ranking.

\paragraph{Feature System.}
Our model retains a feature system aligned with the production baseline, as discarding statistical cross features causes a substantial performance drop that \textit{cannot} be compensated by scaling up the model~\cite{han2025mtgr}. The feature system comprises five categories: (i) \textit{query-document cross statistics} such as exposures and clicks under multiple time windows; (ii) \textit{document-side statistics} such as unique exposures and review counts; (iii) \textit{query-side features} such as query PV and query-document time gaps; (iv) \textit{contextual features} such as display position and exposure time; and (v) \textit{textual features}, i.e., the raw text of queries and documents.

\paragraph{User Behavior Sequence Construction.}
 The core data augmentation over the discriminative baseline is the construction of per-user historical search behavior sequences.
 We aggregate per-user historical search behaviors from online logs and, for each training sample, construct a behavior sequence $S_u = \{(q_1, d_1, t_1), (q_2, d_2, t_2), \ldots\}$. Each historical item carries the complete context of that interaction, including: \textit{query information} (the query text and its categorical taxonomy), \textit{document information} (title, summary, etc.), \textit{interaction signals} (click position, click time, confident-click flag), \textit{historical model scores} (CTR score, AC score, etc.), \textit{multi-dimensional quality signals} (authority, content quality), and \textit{contextual information} (page index, city). To prevent training-time leakage, we strictly filter out behaviors whose timestamps are later than the current request, ensuring that the model only observes ``past'' information. The sequence is sorted by timestamp in ascending order and truncated to a maximum length of $1{,}000$, keeping the most recent behaviors.

\subsection{Framework Overview}
\label{sec:overview}
QGS reformulates search ranking as a query-conditioned generative sequence modeling problem. Given a user's historical search interactions, the framework encodes each interaction as a query-item pair token, models the sequence with a linear-complexity encoder, and predicts the next clicked item conditioned on the current query. As illustrated in Figure~\ref{fig:framework}, QGS consists of four components:
\begin{enumerate}
    \item Query-Item Pair Token Construction (Section~\ref{sec:qi_pair}): each search interaction is encoded as a unified token carrying both query and item information, serving as the input to the sequence encoder.
    \item \textbf{Query-Conditioned Next-Item Prediction} (Section~\ref{sec:infonce}): an InfoNCE contrastive objective that conditions next-item prediction on the current query, which is the central design of QGS.
    \item Linear HSTU Encoder (Section~\ref{sec:linear_hstu}): reformulates HSTU's attention into a linear recurrence, reducing complexity from $O(L^2)$ to $O(L)$ to meet strict online latency budgets.
    \item \textbf{HFG-Attention} (Section~\ref{sec:sparse_attn}): groups heterogeneous sparse contextual features into semantic clusters and fuses them with dense sequential representations via an HSTU attention block.
\end{enumerate}

\subsection{Query-Item Pair Token Construction}
\label{sec:qi_pair}
Generative recommendation models assume that the behavior sequence consists of a single token type---items---because consecutive items typically share topical continuity. In search, however, user behavior is fundamentally different: every interaction is triggered by a query, and each query switch brings an entirely fresh candidate set. Existing generative search models still construct autoregressive sequences from item tokens, treating query information as either a shared prefix condition or a side input. This prevents the sequence from capturing the query-conditioned dependency structure that distinguishes search from recommendation.

To address this issue, we represent each historical search interaction as a unified \textbf{query-item pair token}, which natively encodes the complete information of both the query and the interacted item.

Specifically, given a user's chronologically ordered search interaction sequence $\{(q_1, d_1), (q_2, d_2), \ldots, (q_L, d_L)\}$, for the $t$-th interaction $(q_t, d_t)$, we first concatenate the query text and document text into a single input and feed it into a pre-trained BERT~\cite{devlin2019bert} encoder. We then extract two semantic representations:
\begin{itemize}
    \item \textbf{[CLS] token output} $\mathbf{e}_t^{\text{cls}} \in \mathbb{R}^{d_b}$: serving as the item-level semantic embedding, where $d_b$ denotes the hidden dimension of the BERT encoder.
    \item \textbf{First [SEP] token output} $\mathbf{e}_t^{\text{sep}} \in \mathbb{R}^{d_b}$: serving as the query-level semantic embedding.
\end{itemize}
Meanwhile, we collect a set of numerical and categorical features $\mathbf{f}_t \in \mathbb{R}^{d_f}$ (e.g., click-through rates, dwell-time statistics, matching scores) for the interaction, where $d_f$ denotes the total feature dimension after embedding lookup. The query-item pair token is formed by concatenation:
\begin{equation}
\label{eq:pair_token}
    \mathbf{x}_t = [\mathbf{f}_t \,\|\, \mathbf{e}_t^{\text{cls}}] \in \mathbb{R}^{d_f + d_b}.
\end{equation}
The query semantic embedding $\mathbf{e}_t^{\text{sep}}$ is preserved separately for the prediction head, which is responsible for query-conditioned next-item prediction (see Section~\ref{sec:infonce}). Finally, the concatenated token is projected to the encoder hidden dimension $d_h$ via a linear layer with positional encoding added:
\begin{equation}
    \mathbf{h}_t^{(0)} = W_{\text{proj}} \mathbf{x}_t + \mathbf{p}_t.
\end{equation}

This design offers two advantages. First, unlike methods such as HLLM that use only text semantic embeddings as the SID, our pair token fuses \textbf{query-document text semantics} with \textbf{contextual features} into a unified representation. In industrial search, the same document may elicit vastly different user behaviors at different positions, time periods, or matching qualities. Text semantics alone cannot distinguish these differences. The pair token enriched with contextual features more completely characterizes each interaction; this is validated in our ablation study (Section~\ref{sec:ablation}). Second, encoding each step as a query-item pair makes the sequence faithfully reflect the conditional dependency structure of search behavior. The model can learn that the same user exhibits different interaction patterns under different queries.

\subsection{Query-Conditioned Next-Item Autoregressive Training Objective}
\label{sec:infonce}
The query-item pair token constructed above embeds query information into every sequence position. However, this representation-level change alone is not sufficient. Without explicit conditioning at the training objective level, the model can still predict the next item from historical context alone, ignoring the query signal. In search, the next interacted item depends strongly on the next query. We therefore need a training objective that forces the model to learn $P(\mathrm{item}_{t+1} \mid \mathrm{context}_{\leq t}, \mathrm{query}_{t+1})$ rather than the query-agnostic marginal $P(\mathrm{item}_{t+1} \mid \mathrm{context}_{\leq t})$.

\begin{figure}[t]
  \centering
  \includegraphics[width=\columnwidth]{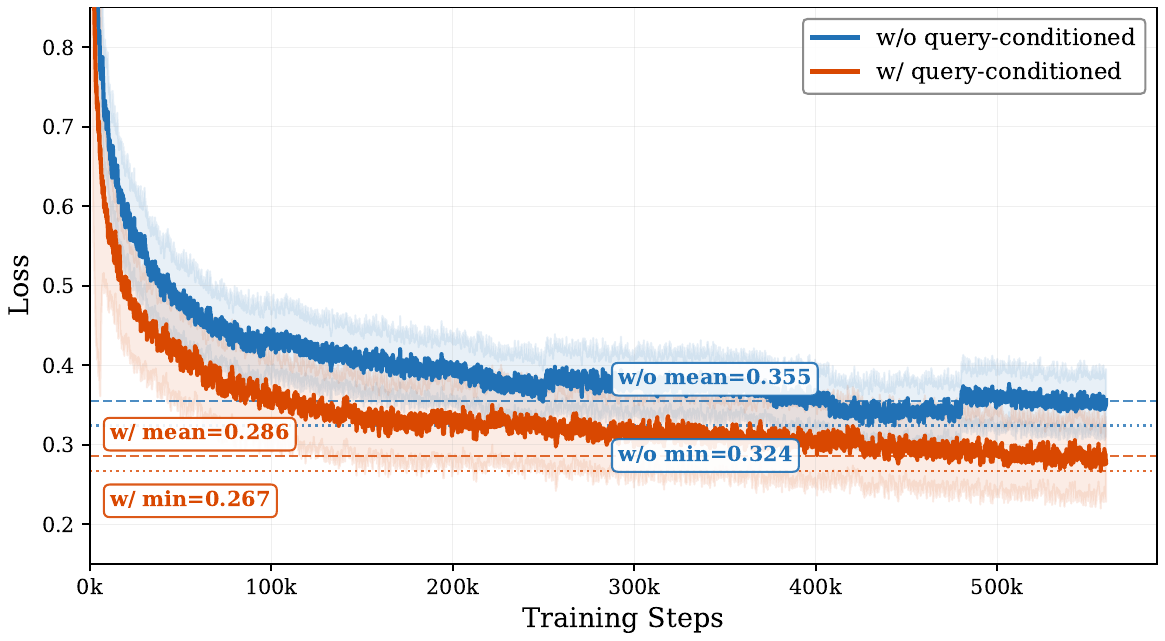}
  \caption{Training loss curves with and without query-conditioned next-item prediction. The item-only objective suffers from high and volatile loss due to noisy supervision caused by query switches, while the query-conditioned objective converges to a significantly lower loss, confirming that the query is a necessary condition for clean supervision in generative search.}
  \label{fig:loss_motivation}
\end{figure}

We empirically validate this motivation. Figure~\ref{fig:loss_motivation} compares the training loss curves of two variants using the same generative architecture with query-item pair tokens as input: one trained with an item-only next-item prediction objective (predicting the next item without the next query), and the other with our query-conditioned objective. The item-only variant exhibits a significantly higher and more volatile loss throughout training, confirming that query switches inject substantial noise into the supervision signal. Once the next query is provided as a condition, the loss drops sharply and converges more smoothly, demonstrating that query conditioning directly resolves the semantic discontinuity inherent in search sequences.

We introduce \textbf{query-conditioned next-item prediction} as the autoregressive training objective. For each position $t$ in the sequence, we concatenate the encoder output at position $t$ with $\mathbf{e}_{t+1}^{\text{sep}}$, the query-level BERT embedding of $q_{t+1}$ (defined in Section~\ref{sec:qi_pair}), to construct the prediction vector:
\begin{equation}
    \mathbf{z}_t = \mathrm{PredictHead}\!\left([\mathbf{h}_t^{(N)} \,\|\, \mathbf{e}_{t+1}^{\text{sep}}]\right).
\end{equation}
The positive target is the projected item representation at position $t+1$: $\mathbf{v}_{t+1} = W_{\text{tgt}} \mathbf{x}_{t+1}$, where $\mathbf{x}_{t+1}$ is the query-item pair token defined in Eq.~\eqref{eq:pair_token}---which encodes both query-document joint semantics (via the BERT [CLS] embedding of the concatenated query-document input) and contextual interaction features---and $W_{\text{tgt}}$ is a learnable projection matrix. Negative samples are drawn from other sequences within the same mini-batch. The similarity matrix is computed via normalized inner product:
\begin{equation}
    s_{t}^{(b, c)} = \frac{\bar{\mathbf{z}}_t^{(b)} \cdot \bar{\mathbf{v}}_{t+1}^{(c)}}{\tau},
\end{equation}
where $\bar{\cdot}$ denotes $L_2$ normalization, the superscripts $(b, c)$ index different sequences within the batch, and $\tau$ is the temperature hyperparameter.

To ensure training quality, we apply two masking strategies on the negative samples:
\begin{itemize}
    \item \textbf{Padding mask.} Negative candidates from padded positions (invalid tokens) are excluded by setting their logits to $-\infty$.
    \item \textbf{Collision mask.} If two sequences happen to contain the same item at the same position, the duplicate is excluded from the negative set to avoid false negatives.
\end{itemize}

The InfoNCE~\cite{oord2018representation} loss is computed as:
\begin{equation}
    \mathcal{L}_{\text{InfoNCE}} = -\frac{1}{|\mathcal{V}|} \sum_{t \in \mathcal{V}} \log \frac{\exp\!\left(s_t^{(b,b)}\right)}{\sum_{c=1}^{B} \exp\!\left(s_t^{(b,c)}\right)},
\end{equation}
where $\mathcal{V}$ is the set of valid (non-padded) positions and $B$ is the batch size.

This design has two key advantages. First, the prediction head \textbf{shares weights} with the final ranking prediction. At inference time, the encoder output at the last position is concatenated with the current query embedding and passed through the same head. This strict alignment prevents train-test mismatch. 

Second, this prevents the model from collapsing to a query-agnostic marginal predictor: prediction must simultaneously rely on historical context and the current query.However, since the next query is used as a prediction condition, a natural concern arises: does this introduce temporal leakage? Our answer is no. \textbf{No temporal leakage.} A natural concern is whether conditioning on $\mathrm{query}_{t+1}$ leaks future information into the historical representation $\mathbf{h}_t^{(N)}$. We emphasize that this is \emph{not} the case. The encoder is strictly causal. The linear recurrence aggregates information via $\mathbf{C}_t = \gamma\,\mathbf{C}_{t-1} + \mathbf{S}_t$ (Eq.~8). Therefore, $\mathbf{h}_t^{(N)}$ depends only on positions $\tau \leq t$. The next-query embedding $\mathbf{e}_{t+1}^{\text{sep}}$ is fused \emph{after} the encoder forward pass completes. It enters only inside the prediction head (Eq.~3) and never participates in any encoder layer computation. At inference time, $\mathrm{query}_{t+1}$ is simply the user's current search query. It is the very input that triggers the ranking request. No privileged future information is required. The design thus maintains strict temporal causality while faithfully modeling the query-driven nature of search.

\begin{figure*}[t]
    \centering
    \includegraphics[width=0.65\linewidth]{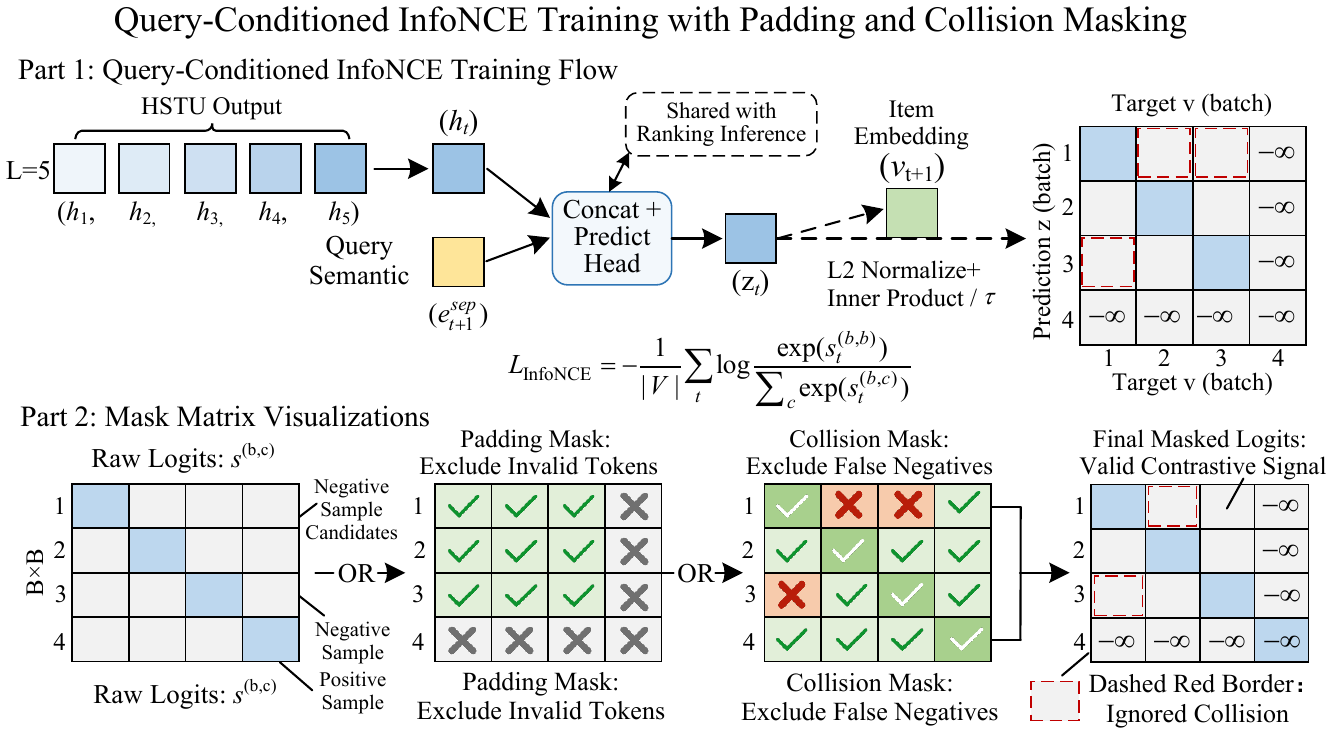}
    \caption{Query-conditioned InfoNCE training with padding and collision masking. \textbf{Part 1} illustrates the training flow: HSTU outputs are concatenated with the next-query semantic embedding, passed through a shared prediction head, $\ell_2$-normalized, and scored against in-batch item embeddings. \textbf{Part 2} visualizes the two masking strategies---the padding mask excludes invalid (padded) tokens, the collision mask excludes false negatives caused by duplicate items across sequences, and their combination yields the final masked logits used for the contrastive objective.}
    \label{fig:infonce}
\end{figure*}

\subsection{Linear HSTU Encoder}
\label{sec:linear_hstu}
The query-conditioned objective above requires a sequence encoder to produce $\mathbf{h}_t^{(N)}$ at every position. Realizing this over up to $1{,}000$ historical interactions, however, is challenging: search ranking must score hundreds of candidates per request within a strict online latency budget. HSTU adopts pointwise aggregated attention that replaces softmax with a SiLU activation and introduces a multiplicative U-gate for output modulation. While this removes the softmax bottleneck, the attention computation still requires an $O(L^2)$ query-key interaction matrix, which is prohibitive for long sequences under such latency constraints.

Inspired by the linear attention paradigm of RWKV~\cite{peng2023rwkv}, we reformulate HSTU's pointwise aggregated attention into a \textbf{linear recurrence} based on element-wise key-value accumulation and causal cumulative summation.

Specifically, for each attention layer, given the input $\mathbf{H} \in \mathbb{R}^{L \times d_h}$, we first apply RMSNorm~\cite{zhang2019root} normalization and then compute four projections:
\begin{equation}
\begin{aligned}
    \mathbf{Q} &= \mathrm{SiLU}(\mathbf{H} W_Q), \quad
    \mathbf{K} = \mathrm{SiLU}(\mathbf{H} W_K), \\
    \mathbf{V} &= \mathrm{SiLU}(\mathbf{H} W_V), \quad
    \mathbf{U} = \mathrm{SiLU}(\mathbf{H} W_U),
\end{aligned}
\end{equation}
where $\mathrm{SiLU}(x) = x \cdot \sigma(x)$ is bounded below, ensuring numerical stability for multi-layer multiplicative operations. Following HSTU's softmax-free design, SiLU bypasses any normalization into a probability distribution.

Instead of computing the full $L \times L$ attention matrix $\mathbf{Q}\mathbf{K}^\top$, we first perform an element-wise multiplication of keys and values:
\begin{equation}
    \mathbf{S}_t = \mathbf{K}_t \odot \mathbf{V}_t \in \mathbb{R}^{d_h}.
\end{equation}
Historical information is then aggregated via causal cumulative summation:
\begin{equation}
    \mathbf{C}_t = \sum_{\tau=1}^{t} \gamma^{t-\tau} \mathbf{S}_\tau = \gamma \, \mathbf{C}_{t-1} + \mathbf{S}_t,
\end{equation}
where $\gamma \in (0,1)$ is a learnable exponential decay factor that down-weights distant interactions, endowing the model with a soft recency bias. The recursive form on the right enables $O(1)$ per-step updates, which is the key to achieving overall $O(L)$ complexity. Meanwhile, it guarantees that position $t$ only attends to positions $\tau \leq t$, naturally satisfying the causal constraint without an explicit mask matrix.
The output at each position is obtained by doubly modulating the accumulated state with the query and the U-gate:
\begin{equation}
    \mathbf{O}_t = \mathbf{Q}_t \odot \mathbf{C}_t \odot \mathbf{U}_t.
\end{equation}

The linear-recurrence formulation drops multi-head attention: all dimensions are processed element-wise inside the cumulative sum. The two modulation vectors $\mathbf{Q}$ and $\mathbf{U}$ therefore play a critical role. $\mathbf{Q}$ acts as \textit{query-side modulation}, selectively extracting relevant information from the accumulated state. $\mathbf{U}$ serves as the \textit{output-side gate}, performing second-stage filtering. They are computed by independent projection matrices and modulate the state from different subspaces, capturing diverse feature-interaction patterns even without explicit multi-head splitting.

Following HSTU, we \textbf{remove the FFN} from each layer. The SiLU activations combined with Q-U dual multiplicative modulation already provide sufficient non-linear capacity. The final layer output is obtained via dropout and a residual connection:
\begin{equation}
    \mathbf{H}^{(\ell+1)} = \mathbf{H}^{(\ell)} + \mathrm{Dropout}(\mathbf{O}).
\end{equation}
The layer structure is illustrated in Figure~\ref{fig:framework}. We stack $N$ layers ($N = 12$ in our experiments) to form the complete Linear HSTU Encoder, and use the output at the last token position as the user's next-item prediction representation (to be fused with the query) $\mathbf{h}_{\text{seq}} \in \mathbb{R}^{d_h}$.

The entire attention computation is thus reduced from $O(L^2 \cdot d_h)$ to $O(L \cdot d_h)$. Unlike generic linear attention methods that discard HSTU's softmax-free design and U-gate, our formulation preserves both: (i) removing softmax avoids the attention-sink problem and permits direct value accumulation; (ii) the Q-U dual gating provides per-position output modulation. At sequence length $1{,}000$, the Linear HSTU Encoder reduces inference latency by $37\%$ with no loss in ranking quality.

\subsection{HFG-Attention}
\label{sec:sparse_attn}
The query-item pair token, the query-conditioned objective, and the Linear HSTU encoder together address user behavior sequence modeling. However, they inherently exclude cross features between the candidate item and the user---since the candidate is the prediction target, not an input. In industrial search ranking, cross features such as query-document matching scores, position-dependent click-through rates, and real-time context signals are critical for ranking quality. Excluding them causes a significant performance drop that cannot be compensated by scaling up the model~\cite{han2025mtgr}. It is therefore necessary to reintroduce contextual and cross features at the top of the model.

These features pose a further challenge: they are heterogeneous. Some are high-dimensional sparse categorical signals; others are low-dimensional numerical values. After embedding lookup, their dimensions and semantic spaces differ substantially. Directly concatenating them with dense sequential representations is ineffective.

We propose Heterogeneous Feature Grouping Attention (HFG-Attention) to bridge this semantic gap. The core idea has three steps: organize heterogeneous features into semantically coherent groups, project each group into a unified embedding space, and apply HSTU attention for adaptive cross-group interaction. We also \textbf{reintroduce the FFN} that HSTU removes, to enhance non-linear fusion capacity across feature groups.

Specifically, we define $G$ feature groups based on semantic relevance. For each group $g$, the constituent sparse features are first converted to dense representations via embedding lookup and concatenated into a group-level vector $\mathbf{r}_g \in \mathbb{R}^{d_g}$, where $d_g$ varies across groups. Each group vector is then projected to a unified target dimension $d_e$ via a linear layer:
\begin{equation}
    \tilde{\mathbf{r}}_g = \varphi(W_g \mathbf{r}_g + \mathbf{b}_g) \in \mathbb{R}^{d_e}.
\end{equation}
To enable the grouping attention to perceive the global feature context, we further project the vector $\mathbf{h}_{\text{dnn}}$ produced by the shared DNN over all raw features to $d_e$ dimensions and append it to the token sequence as a global context token. The resulting structured token sequence is:
\begin{equation}
    \mathbf{R} = [\tilde{\mathbf{r}}_1 \,\|\, \tilde{\mathbf{r}}_2 \,\|\, \cdots \,\|\, \tilde{\mathbf{r}}_G \,\|\, \tilde{\mathbf{h}}_{\text{dnn}}] \in \mathbb{R}^{(G+1) \times d_e}.
\end{equation}

This token sequence is fed into a single-layer HSTU attention block (using standard $O(G^2)$ pointwise aggregated attention since $G$ is small) for adaptive cross-group feature interaction. Unlike the FFN-free design of HSTU, we \textbf{retain the FFN layer} here: because the semantic differences across sparse feature groups are large, the additional non-linear transformation provided by the FFN further fuses heterogeneous features in the unified embedding space. The output is aggregated via masked average pooling and projected to the final fusion dimension:
\begin{equation}
    \mathbf{h}_{\text{sparse}} = W_{\text{out}} \cdot \mathrm{MaskedAvgPool}\!\left(\mathrm{HSTU}(\mathbf{R})\right) \in \mathbb{R}^{d_o},
\end{equation}
where $d_o$ denotes the output fusion dimension.
In the final ranking stage, $\mathbf{h}_{\text{sparse}}$ is concatenated with the candidate document embedding, the shared-DNN-encoded vector $\mathbf{h}_{\text{dnn}}$, and the sequence representation $\mathbf{h}_{\text{seq}}$ produced by the Linear HSTU encoder. The concatenated vector is passed through LayerNorm and forwarded to downstream task towers for fusion prediction.

This design offers two advantages. First, the grouped projection aligns features from disparate semantic spaces into a shared dimension, resolving the semantic gap before interaction. Second, the retained FFN enhances non-linear fusion capacity, enabling richer cross-group feature combinations beyond what attention alone can achieve.

\subsection{Implementation Details}
\label{sec:impl}
\paragraph{Sequence Modeling.}
The maximum length of the user historical search interaction sequence is set to $L_{\max} = 1{,}000$. Each query-item pair token contains $d_f = 151$ dimensions of numerical and categorical features and $d_b = 64$ dimensions of BERT semantic embeddings (CLS and SEP), concatenated and projected to a hidden space of $d_h = 512$ dimensions via a linear layer. We adopt learnable positional encodings. The maximum input length of the BERT encoder is truncated to $180$ tokens.

\paragraph{Query-Conditioned InfoNCE.}
a two-layer MLP with ReLU The temperature hyperparameter is set to $\tau = 0.1$.

\paragraph{Linear HSTU Encoder.}
The encoder stacks $N = 12$ layers with hidden dimension $d_h = 512$.

\paragraph{HFG-Attention.}
After embedding lookup, each feature group is projected to a unified dimension of $d_e = 16$ via a fully connected layer. The output of the shared DNN is passed through a stop-gradient operation and similarly projected to $d_e$ dimensions, then appended as a global context token. We use a single-layer standard HSTU attention ($O(G^2)$ complexity, $8$ attention heads) with the FFN layer retained (intermediate dimension $4 \times d_e = 64$). The output is aggregated via masked average pooling and projected to $128$ dimensions.

\section{Experiments}

\subsection{Experimental Setup}
\label{sec:exp_setup}
We design experiments to answer four research questions:
\textbf{RQ1}: Does QGS outperform strong baselines? (Section~\ref{sec:overall})
\textbf{RQ2}: Which modules contribute the gain? (Section~\ref{sec:ablation})
\textbf{RQ3}: How does QGS scale with depth, width, and sequence length? (Section~\ref{sec:scalability})
\textbf{RQ4}: Do offline gains transfer to online metrics? (Section~\ref{sec:online})

\paragraph{Datasets.}
We construct training data from production logs of Quark Search rather than public benchmarks.
There are four reasons for this choice.
First, our model relies on long-term user behavior sequences and real-time session-level interaction signals such as clicks, dwell time, and conversions. Public datasets rarely provide such sequential depth.
Second, production ranking requires comprehensive contextual features, fine-grained user--item cross features, and statistical features such as historical CTR and category preferences. Most public datasets contain only query and document text. They lack these side information.
Third, our dataset comprises tens of millions of users and hundreds of millions of impressions. Public search corpora are typically orders of magnitude smaller. They cannot support the data-hungry training of generative ranking models.
Fourth, our core conclusions depend on online A/B testing to measure real business metrics. Public datasets only permit offline evaluation. They cannot validate end-to-end deployment effectiveness.

For offline evaluation, we collect $18$ days of search logs, yielding 488M impression records (Table~\ref{tab:dataset}). For online evaluation, we train on more than one month of production logs and compare against the production baseline via A/B testing.

\begin{table}[t]
    \centering
    \caption{Statistics of the offline training and test sets collected from Quark Search.}
    \label{tab:dataset}
    \begin{tabular}{lrrrr}
        \toprule
        Split & \#Users & \#Searches & \#Impressions & \#Clicks \\
        \midrule
        Train & 37.87M & 154.77M & 488.58M & 82.58M \\
        Test  &  2.10M &   8.60M &  27.14M &  4.59M \\
        \bottomrule
    \end{tabular}
\end{table}

\paragraph{Baselines.}
We compare the proposed method against three baselines that span the discriminative paradigm and two representative generative sequence-modeling paradigms:
\begin{itemize}
    \item \textbf{Production Baseline (Base).} The currently deployed discriminative deep ranking model, built on a BERT\&DNN architecture combined with hand-crafted feature engineering. It represents the mature industrial paradigm for search ranking.
    \item \textbf{HLLM~\cite{Chen2024HLLMES}.} Both the Item LLM and User LLM are initialized from a pre-trained BERT and configured as $12$-layer Transformers with $12$ attention heads and a hidden size of $768$. Each token in the sequence carries only the textual semantic embedding as its SID, with no contextual features included.
    \item \textbf{HSTU~\cite{Zhai2024ActionsSL}.} The standard pointwise aggregated attention configured with $12$ layers, $8$ attention heads, and a hidden size of $768$.
\end{itemize}
We exclude OneSearch, UniSearch, and GenSAR because their formulations target retrieval-oriented generation or unified search--recommendation settings, which do not fit our industrial ranking scenario with strict latency constraints and rich cross features. HSTU and HLLM serve as general autoregressive backbones that can be adapted to the same ranking protocol, offering a more controlled comparison.

\paragraph{Evaluation Metrics.}
\emph{Offline.} We report \textbf{AUC} and \textbf{GAUC} for click-through rate prediction. GAUC averages AUC per search request and better reflects within-session ranking quality.

\emph{Online.} We report three metrics: (i) \textbf{CTR}---overall click-through rate; (ii) \textbf{Click-Search Ratio}---proportion of searches with at least one click; (iii) \textbf{PV Duration}---average dwell time per page view.

\paragraph{Implementation Details.}
All models are trained with the Adagrad optimizer, a learning rate of $0.01$, and a batch size of $100$, on $8$ NVIDIA A800 GPUs. For the comparison experiments, the input sequence length is set to $100$ uniformly across all baselines and our method to ensure a fair comparison.

\subsection{Overall Performance Comparison}
\label{sec:overall}
To answer \textbf{RQ1}, we compare all models on the offline test set (Table~\ref{tab:overall}).

\begin{table}[t]
    \centering
    \caption{Offline comparison on the Quark Search test set. Best results are in \textbf{bold}. Higher is better for both metrics ($\uparrow$). \colorbox{bestcolor}{Orange} = top-1; \colorbox{secondcolor}{Yellow} = top-2.}
    \label{tab:overall}
    \begin{tabular}{lcc}
        \toprule
        Model & GAUC $\uparrow$ & AUC $\uparrow$ \\
        \midrule
        Base (Bert\&DNN) & 0.7140 & 0.8040 \\
        HSTU~\cite{Zhai2024ActionsSL} & 0.7412 & 0.8371 \\
        HLLM~\cite{Chen2024HLLMES}    & \cellcolor{secondcolor}0.7436 & \cellcolor{secondcolor}0.8394 \\
        \midrule
        \textbf{QGS (Ours)} & \cellcolor{bestcolor}\textbf{0.7573} & \cellcolor{bestcolor}\textbf{0.8541} \\
        \midrule
        Impr.\ over Base (\%)       & \textbf{6.06}   & \textbf{6.23}   \\
        \bottomrule
    \end{tabular}
\end{table}

\textbf{(1) QGS significantly outperforms the production baseline.} GAUC improves by $+4.33$ absolute points ($0.7140 \rightarrow 0.7573$) and AUC by $+5.01$ points ($0.8040 \rightarrow 0.8541$). This validates the advantage of generative query-item pair sequence modeling over the discriminative paradigm.

\textbf{(2) QGS also outperforms HSTU and HLLM.} Over HSTU, QGS gains $+1.61$ GAUC and $+1.70$ AUC. Over the strongest generative baseline HLLM, the gains are $+1.37$ GAUC and $+1.47$ AUC. Explicitly modeling query-item conditional dependency brings additional gains beyond what the encoder backbone alone can offer.

\textbf{Takeaway.} The query-conditioned generative formulation consistently outperforms both the discriminative production system and strong item-only generative baselines, confirming that search ranking benefits from explicitly coupling queries and items in the autoregressive sequence.

\subsection{Ablation Studies}
\label{sec:ablation}
To answer \textbf{RQ2}, we ablate QGS by removing exactly one component at a time. Results are in Table~\ref{tab:ablation}.

\begin{table}[t]
    \centering
    \caption{Ablation study on the offline test set. \textit{Inference} reports the relative latency change w.r.t.\ the full model (positive numbers indicate slowdown). Higher is better for GAUC/AUC ($\uparrow$).}
    \label{tab:ablation}
    \small
    \setlength{\tabcolsep}{3pt}
    \begin{tabular}{@{}lccc@{}}
        \toprule
        Variant & GAUC $\uparrow$ & AUC $\uparrow$ & Inference \\
        \midrule
        \textbf{QGS (full)} & \cellcolor{secondcolor}\textbf{0.7573} & \cellcolor{secondcolor}\textbf{0.8541} & baseline \\
        \midrule
        w/o Feature Grouping Attn.          & 0.7410          & 0.8433          & $+3\%$    \\
        w/o Query-cond.\ next-item pred.    & 0.7382          & 0.8421          & baseline  \\
        w/o Contextual feats.\ in pair tok. & 0.7549          & 0.8476          & baseline  \\
        Linear HSTU $\to$ original HSTU     & \cellcolor{bestcolor}\textbf{0.7575} & 0.8540          & $+37\%$   \\
        Linear HSTU $\to$ pre-trained LLM   & 0.7572          & \cellcolor{bestcolor}\textbf{0.8549} & $+61\%$   \\
        \bottomrule
    \end{tabular}
\end{table}

\textbf{(1) Query-conditioned next-item prediction is the most impactful design.} This variant removes only the next-query condition from the prediction head; the InfoNCE loss is still used. GAUC drops by $1.91$ points and AUC by $1.20$. In search, predicting the next item from historical context alone is insufficient. The query condition provides a supervision signal tightly aligned with the ranking objective.

\textbf{(2) HFG-Attention is the second most critical module.} Removing it degrades GAUC by $1.63$ and AUC by $1.08$. This confirms that grouping attention is essential for bridging the semantic gap between heterogeneous sparse features and dense sequential representations.

\textbf{(3) Contextual features in the pair token are non-trivial.} Removing them costs $0.24$ GAUC and $0.65$ AUC. Text semantics alone cannot capture interaction differences under varying contexts (position, time, traffic source).

\textbf{(4) Linear HSTU achieves the best efficiency-quality trade-off.} Replacing it with the original HSTU leaves metrics unchanged but adds $37\%$ latency. Replacing it with a pre-trained LLM yields marginal AUC gain at $61\%$ higher latency. Linear HSTU matches heavyweight encoders in quality while substantially reducing inference cost.

\subsection{Scalability}
\label{sec:scalability}
To answer \textbf{RQ3}, we vary encoder depth, hidden dimension, and sequence length (Table~\ref{tab:scalability}).

\begin{table}[t]
    \centering
    \caption{Scalability of the proposed method along encoder depth, hidden dimension, and sequence length. Default configuration is $12$L, $512$d, $100$len.}
    \label{tab:scalability}
    \begin{tabular}{lcc}
        \toprule
        Configuration & GAUC $\uparrow$ & AUC $\uparrow$ \\
        \midrule
        $12$L, $512$d, $100$len  & \cellcolor{secondcolor}0.7573 & \cellcolor{secondcolor}0.8541 \\
        \midrule
        $1$L,  $512$d, $100$len  & 0.7478          & 0.8464         \\
        $4$L,  $512$d, $100$len  & 0.7493         & 0.8503         \\
        $12$L, $128$d, $100$len  & 0.7516          & 0.8485       \\
        $12$L, $512$d, $1000$len & \cellcolor{bestcolor}\textbf{0.7604} & \cellcolor{bestcolor}\textbf{0.8567} \\
        \bottomrule
    \end{tabular}
\end{table}

\textbf{Longer sequences bring consistent gains.} Scaling from $100$ to $1{,}000$ tokens improves GAUC by $+0.31$ ($0.7573 \rightarrow 0.7604$) and AUC by $+0.26$ ($0.8541 \rightarrow 0.8567$). The Linear HSTU encoder benefits from longer behavior contexts.

\textbf{Depth and width both contribute meaningfully.} Scaling depth from $4$ to $12$ layers gains $+0.80$ GAUC and $+0.38$ AUC, substantially larger than the $1$-to-$4$ layer gain ($+0.15$ GAUC), indicating that deeper stacking continues to improve representation quality. Widening the hidden dimension from $128$ to $512$ at full depth gains $+0.57$ GAUC and $+0.56$ AUC. Under the linear-attention design, all three scaling axes---depth, width, and sequence length---remain effective, with longer sequences providing the most consistent returns.

\subsection{Online A/B Experiment}
\label{sec:online}
To answer \textbf{RQ4}, we deploy QGS in the CTR prediction module of Quark Search and run an online A/B test against the production baseline on 2\% of traffic for 7 consecutive days. The experimental traffic covers millions of exposures per day. All reported lifts are statistically significant ($p < 0.05$).

\begin{table}[t]
    \centering
    \caption{Online A/B test results on Quark Search. \textit{Lift} is the relative improvement over the production baseline.}
    \label{tab:online}
    \begin{tabular}{lc}
        \toprule
        Metric & Lift over Base \\
        \midrule
        CTR $\uparrow$                            & +0.62\% \\
        Click-Search Ratio $\uparrow$             & +0.38\% \\
        PV Duration $\uparrow$                    & +3.55\% \\
        \bottomrule
    \end{tabular}
\end{table}

As shown in Table~\ref{tab:online}, QGS improves CTR by \textbf{+0.62\%}, Click-Search Ratio by \textbf{+0.38\%}, and PV Duration by \textbf{+3.55\%} over the production baseline. The PV Duration gain indicates deeper user engagement with ranked results. These results confirm that offline gains transfer consistently to the online setting.

\section{Conclusion}

In this paper, we propose \textbf{QGS}, a query-conditioned generative search framework that bridges the structural mismatch between recommendation-style generative sequence modeling and industrial search ranking. By representing each user history step as a query-item pair token and supervising the model with a query-conditioned next-item objective, QGS replaces the noisy marginal next-item distribution with a clean conditional one, eliminating the sequence-level discontinuity caused by query switches. To make this design deployable, we further introduce a Linear HSTU encoder that reduces per-layer complexity from $O(L^2)$ to $O(L)$, and an HFG-Attention block that bridges the semantic gap between sequential embeddings and heterogeneous sparse cross features. Offline experiments on Quark Search show that QGS consistently outperforms strong discriminative and generative baselines, and online A/B testing confirms its effectiveness in production.

\noindent\textbf{Limitations and Future Work.} The current framework focuses on group-level semantic fusion for cross-feature modeling and has not yet explored finer-grained token-level feature interaction paradigms such as RankMixer~\cite{zhu2025rankmixer}. In addition, our evaluation is limited to a fixed-length recent behavior window, and the effectiveness of QGS on ultra-long user lifetime histories remains to be validated. We plan to investigate these directions and to extend QGS toward unified multi-scenario search-and-recommendation modeling.

\section*{Declaration of Generative AI and AI-assisted technologies in the writing process}

During the writing process of this work, the authors used ChatGPT in order to improve language. After using this tool, the authors reviewed and edited the content as needed and take full responsibility for the content of the publication.

\clearpage
\bibliographystyle{ACM-Reference-Format}
\bibliography{references}

\end{document}